\documentclass[10pt,letterpaper]{article}

\usepackage[utf8]{inputenc}

\usepackage{cite}

\usepackage{nameref,hyperref}

\usepackage{microtype}
\DisableLigatures[f]{encoding = *, family = * }

\raggedright
\usepackage{changepage}

\usepackage[aboveskip=1pt,labelfont=bf,labelsep=period,singlelinecheck=off]{caption}

\usepackage{lastpage,fancyhdr,graphicx}
\usepackage{epstopdf}
\fancyheadoffset[L]{2.25in}
\fancyfootoffset[L]{2.25in}

\usepackage{color}

\definecolor{Gray}{gray}{.25}

\usepackage{graphicx}

\usepackage{sidecap}

\usepackage{wrapfig}
\usepackage[pscoord]{eso-pic}
\usepackage[fulladjust]{marginnote}
\reversemarginpar

\begin{document}
\vspace*{0.35in}

\begin{flushleft}
{\Large
\textbf\newline{Single-Cycle Self-Compressed Near Infrared Pulses using High-Spatial Modes in Hollow-Core Fibers.}
}
\newline
\\
Boris A. L\'opez-Zubieta\textsuperscript{1,*},
Enrique Conejero Jarque\textsuperscript{1},
\'I\~nigo J. Sola\textsuperscript{1},
Julio San Roman\textsuperscript{1},
\\
\bigskip
\bf{1} Grupo de Investigaci\'on en Aplicaciones del L\'aser y Fot\'onica, Departamento de F\'isica Aplicada, University of Salamanca, E-37008, Salamanca, Spain
\\
\bigskip
* boris.lopez@usal.es

\end{flushleft}

\section*{Abstract}
Soliton self-compression is demonstrated during the propagation of high spatial modes in hollow core fibers in the near-infrared spectral region, taking advantage of their negative dispersion response. We have found that there is always an optimum spatial mode to observe this phenomenon, compressing the pulses down to the single-cycle regime without needing any external compression device and with a consequent increase in the output peak power. Our result is relevant for any ultrashort laser application in which few- or single-cycle pulses are crucial.


\section*{Introduction}
Ultrashort pulses are nowadays indispensable tools in many scientific disciplines such as chemistry \cite{zewail00}, material science, biology \cite{king05} or physics \cite{corkum07}. The wide range of applications related to these sources forced a rapid development of techniques to achieve shorter and more energetic pulses. The nonlinear propagation techniques proposed in the early 80s to compress laser pulses using optical fibers\cite{mollenauer80} have been adapted and improved to obtain pulses with better properties.

At the mid-nineties, a completely new scenario to explore the nonlinear interaction between light and matter in the high energy regime appeared: the hollow core fibers (HCFs). These glass capillary fibers, filled with a gas, support different linearly polarized leaky modes, called hybrid modes HE$_{1m} (m=1,2,...)$, with well-known dispersion and absorption coefficients \cite{marcatilli64}. The propagation properties of these modes could be tuned by changing the size of the core and/or the pressure of the filling gas, which made them very attractive. The fundamental leaky mode (HE$_{11}$), presenting the lowest absorption, is the most attractive mode to study nonlinear propagation phenomena. It was first used to generate ultra-short pulses in the mJ regime, employing a phase compensation system to eliminate the spectral phase acquired during the nonlinear propagation process\cite{nisoli96,nisoli97}. This post-compression set-up has become the most standard compression technique for femtosecond pulses at the millijoule level with impressive results\cite{paco17}.

Immediately after the appearance of the HCFs, the hollow core PCFs (HC-PCF) were developed\cite{cregan99}. This new type of structured fibers decreases the losses and keeps all the good tuning properties of standard HCFs. They have also been used to generate ultrashort pulses in the $\mu$J level\cite{heckl11}, reaching in some cases durations below the single optical cycle\cite{travers11}. One can also take advantage of the dispersion tunability of these fibers to achieve self-compression of the input pulses by soliton-effect (using the anomalous dispersion response)\cite{ouzounov03}.

Although the fundamental mode is the one usually employed for applications like post-compression, there are several studies that show how to excite higher spatial modes in HCFs\cite{daria04}, metallic hollow core waveguides\cite{Yirmiyahu07} and HC-PCFs\cite{Ishaaya08,euser08}. These studies are focused on generating new laser beams that might be useful for applications like particle guiding, microscopy or laser processing. In this paper we investigate the propagation properties of the excited spatial modes of a HCF. We identify the anomalous dispersion response that some of these excited spatial modes present at atmospheric pressure for near-infrared (NIR) wavelengths. This fact, together with the high nonlinear interaction due to their small spatial dimensions, make them very attractive for self-compression schemes. We have demonstrated that using different gases we are able to obtain single-cycle output pulses from a standard post-compression setup.

\section{Linear response of the leaky modes in the NIR spectral region}
\label{sec:examples}

\begin{figure}[htbp]
\centering
\includegraphics[width=\linewidth]{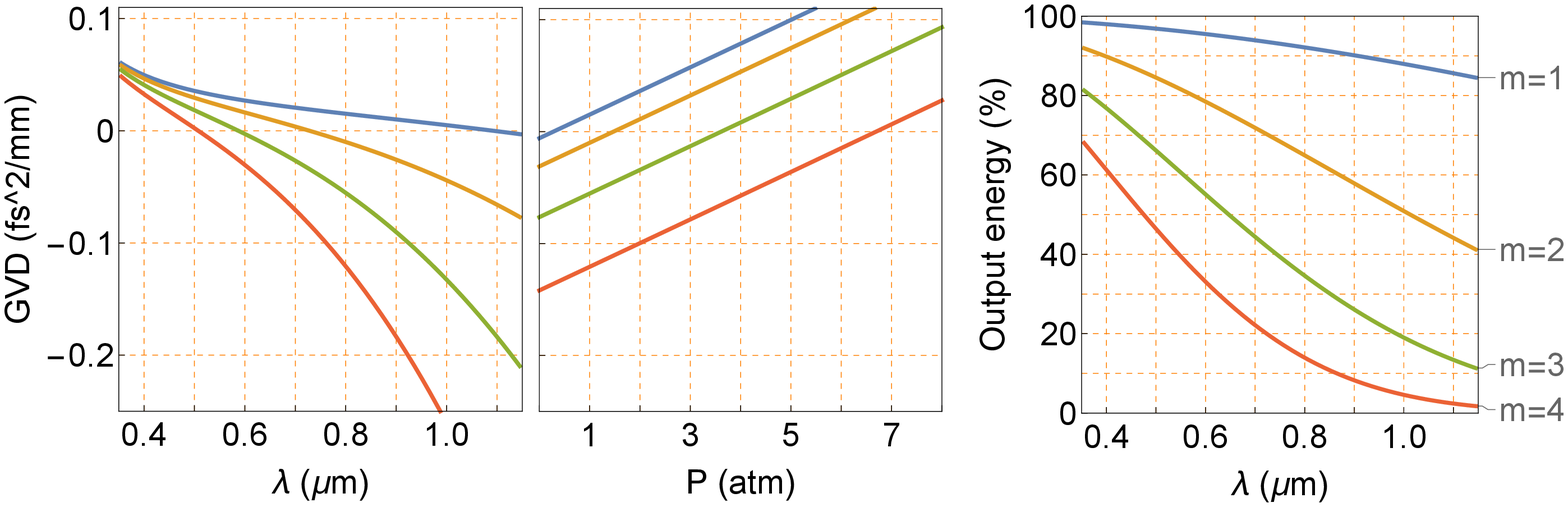}
\caption{GVD of the four lowest spatial modes of a HCF depending on $\lambda$, for Ar at 1 bar, (left) and on pressure (P), at 800 nm, (middle). On the right, we show the wavelength dependence of the throughput energy for a 1 meter long HCF filled with 1 bar of Ar. The HCF has a core radius of 150 $\mu$m in all cases. The blue line represents the fundamental mode ($m=1$), orange the first mode ($m=2$), green the second mode ($m=3$) and red the third mode ($m=4$).}
\label{fig:dispersion}
\end{figure}

The propagation, $\beta_m (\omega)$, and the intensity absorption, $\alpha_m (\omega)$, coefficients of a linearly polarized $m^{th}$-spatial mode of a HCF (the hybrid mode HE$_{1m}$) are \cite{marcatilli64}:  
\begin{equation}
\beta_m ( \omega ) = \frac{n ( \omega ) \omega}{c} \left( 1 + \left( \frac{u_m c}{r_F n ( \omega ) \omega} \right)^2 \right), ~~~m=1,2,...
\label{eq:prop_coef}
\end{equation}
\begin{equation}
\alpha_m ( \omega ) = \left( \frac{u_m}{2\pi} \right)^2 \left( \frac{2\pi c}{n( \omega )\omega} \right)^2 \frac{1}{r_F^3} \frac{\nu^2+1}{\sqrt{\nu^2-1}} ~~~;~~~ \nu = \frac{n_{cl} ( \omega )}{n( \omega )},
\label{eq:abs_coef}
\end{equation}
where $m$ indicates the spatial mode order ($m=1$ representing the fundamental mode), $c$ is the speed of light in vacuum, $u_m$ is the $m^{th}$-zero of the $J_0 (x)$ Bessel function, $r_F$ is the HCF core radius, and $n(\omega)$ and $n_{cl}(\omega)$ are the refractive index of the gas filling the HCF and of the fiber material, respectively. The left and middle pictures of Fig. \ref{fig:dispersion} show the Group Velocity Dispersion (GVD), $( d^2 \beta_m / d \omega^2 )$, as a function of the wavelength (left) and of the gas pressure (middle), for the four lowest spatial modes. All cases correspond to a 150 $\mu$m core radius HCF filled with Ar.

The wavelength dependence of the GVD for an HCF with 150 $\mu$m core radius, filled with 1 bar of Ar (the left picture of Fig. \ref{fig:dispersion}) shows that all the spatial modes except the fundamental present negative dispersion at 800 nm. Therefore, any of them would be a good candidate to achieve self-compression, which comes from the balance between the dispersion and the self-phase modulation (SPM), similarly to what has been done in fibers \cite{mollenauer83} and PCFs \cite{ouzounov05}. The pressure dependence of the GVD for 800 nm (middle picture of Fig. \ref{fig:dispersion}) shows that the modal dispersion prevails over the gas dispersion in the low-pressure regime, all modes presenting an anomalous dispersion response below 0.28 bars. In summary, Fig. \ref{fig:dispersion} indicates that there is plenty of room to take advantage of the anomalous dispersion of the high spatial modes of an HCF.  

To complete the discussion about the usefulness of the high-spatial modes of an HCF we have to take into account their losses. By simple inspection of the intensity absorption coefficient, Eq. \ref{eq:abs_coef}, it is clear that the longer the wavelength, the smaller the radius of the HCF, and/or the highest the spatial mode, the highest the absorption. The right picture of Fig. \ref{fig:dispersion} shows the wavelength dependence of the output energy obtained from the first four spatial modes with a 1 meter long HCF with 150 $\mu$m core radius filled with 1 bar of Ar. It can be seen that at 800 nm the two or three first spatial modes suffer small or moderate losses and can still be used for self-compression applications.

The same conclusions are obtained when using other gases such as Ne or air. We should only remark that the prevalence of the anomalous dispersion region for Ne is maintained over a larger pressure range as compared to Ar or air, as it is the least dispersive gas among them. Apart from this detail, there is no difference in the general tendencies presented here for Ar.

\section{Nonlinear propagation of high spatial modes: looking for the self-compression regime for different gases}

To study the self-compression dynamics in HCF in the NIR spectral region we have simulated the nonlinear propagation of high-spatial modes in a standard HCF post-compression scheme. For these calculations we have used a numerical code that includes all the dispersion terms, the linear absorption, SPM, ionization, self-steepening (SS) and the same high-order nonlinear correction on the ionization term. The linear propagation part is solved by projecting the beam into the first 30 spatial modes, each of them evolving with their own propagation coefficient, as Eqs. \ref{eq:prop_coef} and \ref{eq:abs_coef} indicate. The nonlinear terms are solved using a standard fourth-order Runge-Kutta scheme. For more details of the numerical model see \cite{conejero17}.

\begin{figure}[htbp]
\centering
\includegraphics[width=\linewidth]{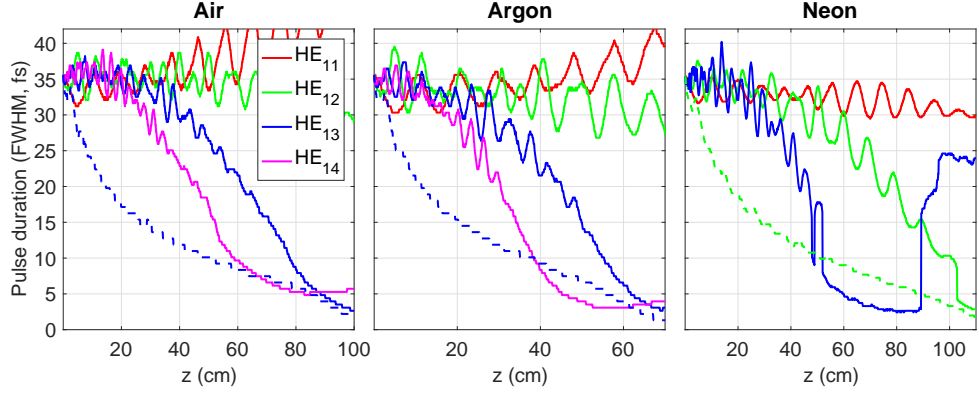}
\caption{Evolution of the pulse duration in a HCF filled with 1 bar of air (left), Ar (middle) and Ne (right). The HCF is 1 m (left), 70 cm (middle) and 110 cm (right) long, always with a core radius of 150 $\mu$m. The red, green, blue and magenta continuous lines represent the fundamental, second, third and fourth spatial mode. The dashed lines represent the FWHM of the Fourier Limit of the optimum case, identified with the same color code. The input pulse has 35 fs FWHM with 80 $\mu$J (air), 75 $\mu$J (Ar) and 0.7 mJ (Ne) of input energy.}
\label{fig:gvd}
\end{figure}
Figure \ref{fig:gvd} shows the temporal duration evolution of the first spatial modes in an HCF filled with air (left), Ar (middle) and Ne (right). All cases correspond to an input laser pulse of 35 fs temporal full width at half maximum (FWHM) at 800 nm. The red, green, blue and magenta continuous lines represent the fundamental (HE$_{11}$), second (HE$_{12}$), third (HE$_{13}$) and fourth (HE$_{14}$) spatial modes, respectively. Note that in the case of Ne (right plot) we have only simulated the first three modes because they suffice to identify the best spatial mode to observe self-compression. Finally, we have added a dashed line with the FWHM of the Fourier Limit of the optimum case, using the same color code to indicate which is the best spatial mode in each case.

Note, first of all, that the temporal durations of any mode of Fig. \ref{fig:gvd} show oscillations related to the interference between the different modes that appear during the nonlinear propagation. These interferences affect the temporal and spatial dimensions of the beam, causing this oscillatory behavior\cite{nurhuda03}.

The left picture of Fig. \ref{fig:gvd} shows the FWHM evolution of four 80 $\mu$J different beams coupled into a 1 m long HCF with 150 $\mu$m core radius filled with 1 bar of air. The fundamental (red line) and the second spatial mode (green line) do not show self-compression, although both cases present spectral broadening during their propagation. In these cases, there is no compensation between negative dispersion and SPM\cite{agrawal}. By contrast, when using the third (blue line) and fourth (magenta line) spatial modes, the pulses self-compress.  See that the fourth spatial mode, which presents a more negative linear dispersion response, self-compresses earlier but to a longer pulse. This is a direct consequence of the higher losses of the higher spatial modes, showing that absorption prevents further self-compression. The third spatial mode (HE$_{13}$) is, therefore, the optimum case for the parameters used in Fig. \ref{fig:gvd}. To have an idea of the quality of the self-compression process we have first added a blue dashed line that represents the evolution of the duration of the Fourier Limited pulse of the optimum mode. We have also calculated the ratio between the energy that remains in the main peak of the Fourier Limited pulse compared to the energy of the main peak of the self-compressed pulse, both at the end of the HCF, $Q_{SC} \sim \Delta\tau^{FL} I_{max}^{FL}/ \Delta\tau I_{max}$, where $\Delta\tau$ represents the FWHM pulse duration and $I_{max}$ is the maximum intensity on axis. In this case the pulse self-compresses to 2.6 fs FWHM, with $Q_{SC}=0.8$ and showing an increase of the peak power from 2.1 GW input peak power to 3.8 GW (1.8 increase factor).

The middle picture of Fig. \ref{fig:gvd} represents the evolution of 75 $\mu$J pulses with different spatial modes in a 70 cm long HCF with 150 $\mu$m core radius filled with 1 bar of argon. The observed dynamics remains basically the same as in air. The optimum self-compression is achieved when the input beam is coupled into the third spatial mode. In this case, the pulse self-compresses to 2.8 fs, with $Q_{SC}=0.8$ and with an increase of the peak power from 2.0 GW to 5.2 GW (2.6 increase factor).

\begin{figure}[htbp]
\centering
\includegraphics[width=12.5 cm]{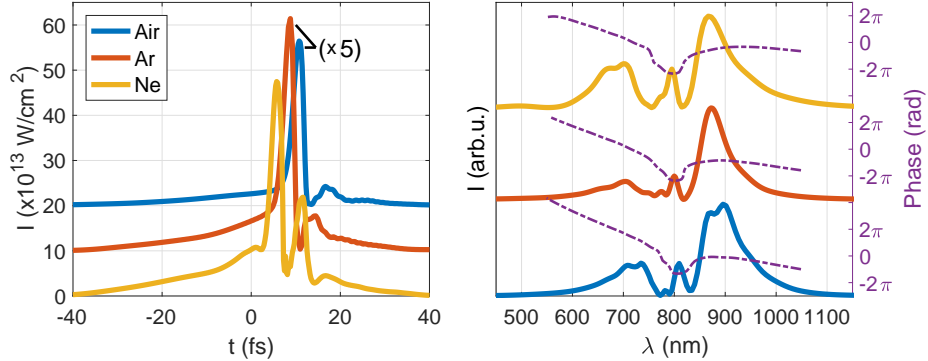}
\caption{Temporal (left) and spectral (right) intensity structure of the output pulse obtained from a HCF filled with air (blue), argon (red) and neon (yellow) under optimal self-compression conditions shown in Fig. \ref{fig:gvd}. The temporal intensity distribution for air and argon have been up-shifted and amplified for comparison purposes. The spectra are accompanied by the spectral phase for each case to have an idea of the compression quality.}
\label{fig:pulses}
\end{figure}
The right picture of Fig. \ref{fig:gvd} shows the case of 0.7 mJ pulses with different spatial modes in a 110 cm long HCF with 150 $\mu$m core radius filled with 1 bar of neon. In this case we identify two optimum output pulses. The second spatial mode (green line) at the end of the HCF or the third spatial mode (blue line) at 85 cm. The second and third spatial mode self-compress to 2.9 fs and 2.6 fs, with $Q_{SC}=0.5$ and $0.4$, with an increase in the peak power from 21.1 GW to 46.1 GW (2.2 increase factor) and 27.6 GW (1.3 increase factor), respectively. Clearly, the second spatial mode is optimum in efficiency, as it has fewer losses, while the third one is better in temporal compression.

Up to now, we have demonstrated that high order spatial modes might present soliton-like self-compression dynamics in different gases. To complete the discussion on the self-compression process we present the temporal (left) and spectral (right) on-axis structure of the pulse at the end of the HCF for air (blue), argon (red) and neon (yellow) in Fig. \ref{fig:pulses}. It shows that the three different gases generate similar pulse structures: from the temporal point of view the pulses present a long temporal pedestal at the front part followed by the very short main peak and some tail structure, indicating the presence of positive third order dispersion (TOD). From the spectral point of view, we observe that the spectra are quite modulated, indicating the relevance of the SPM as the main broadening effect. 

\begin{figure}[htbp]
\centering\includegraphics[width=12 cm]{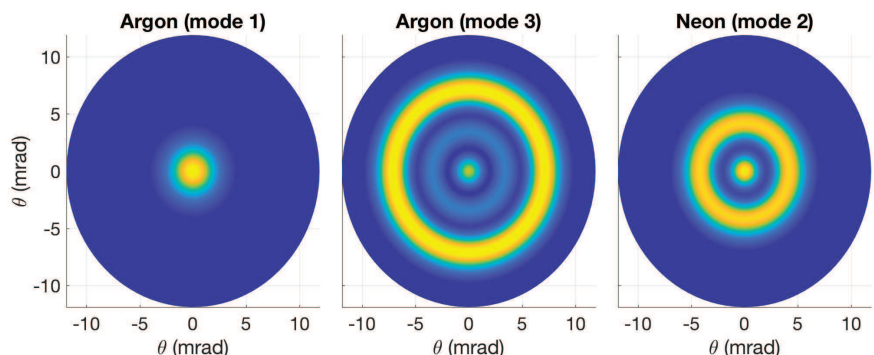}
\caption{Far field distribution of the output beam obtained when propagating the fundamental (left) and third spatial mode (middle) in Ar, or the second spatial mode (right) in Ne. The parameters for each case are the same as in Fig. \ref{fig:gvd}.}
\label{fig:far-field}
\end{figure}

We have shown above that self-compression with NIR pulses occurs in a HCF when coupling the input beam into higher spatial modes. Looking for an easy way to find out if we really succeed in coupling the input beam into some high spatial mode, we have identified that the spatial far-field distribution could be a very intuitive fingerprint of the solitonic-type self-compression that we are discussing. Fig. \ref{fig:far-field} shows the far-field distribution of the output beam when starting the propagation with the first (fundamental) or the third spatial mode in Ar (left and middle picture, respectively), or the second spatial mode in Ne (right picture), using the parameters of Fig. \ref{fig:gvd}. It is evident that the presence of a ring structure in the far field will be an indication that high spatial modes have been excited.  

\section{Conclusion}

We have demonstrated that solitonic-like self-compression can occur in the NIR spectral regime during the propagation of high spatial modes of HCF. We have shown that there is always an optimum spatial mode to observe this phenomenon and, although the losses associated to those modes are important, we have proven that there is always an increase of the output peak power due to the high temporal compression achieved. This behavior can be observed in the NIR spectral region for a wide range of parameters (type of gas, pressure, laser wavelength) thanks to the good control of the dispersion that can be achieved in HCF. 

\section*{Acknowledgments}
Junta de Castilla y Le\'on (SA046U16); MINECO (FIS2013-44174-P, FIS2015-71933-REDT, FIS2016-75652-P); Fundaci\'on Carolina fellowship.




\end{document}